\makeatletter
\newcommand{\dontusepackage}[2][]{%
  \@namedef{ver@#2.sty}{9999/12/31}%
  \@namedef{opt@#2.sty}{#1}}
\makeatother
\dontusepackage{subfigure}

\documentclass[english,]{sigchi-ext}

\usepackage{lmodern}
\usepackage{amssymb,amsmath}
\usepackage{ifxetex,ifluatex}
\usepackage{fixltx2e} 
\usepackage{ragged2e} 
\ifnum 0\ifxetex 1\fi\ifluatex 1\fi=0 
  \usepackage[T1]{fontenc}
  \usepackage[utf8]{inputenc}
\else 
  \ifxetex
    \usepackage{mathspec}
    \usepackage{xltxtra,xunicode}
  \else
    \usepackage{fontspec}
  \fi
  \defaultfontfeatures{Mapping=tex-text,Scale=MatchLowercase}
  
\fi
\IfFileExists{upquote.sty}{\usepackage{upquote}}{}
\IfFileExists{microtype.sty}{%
\usepackage{microtype}
\UseMicrotypeSet[protrusion]{basicmath} 
}{}
\ifxetex
  \usepackage{polyglossia}
  \setmainlanguage{english}
\else
  \usepackage[english]{babel}
\fi
\usepackage{listings}
\usepackage{graphicx}
\makeatletter
\def\maxwidth{\ifdim\Gin@nat@width>\linewidth\linewidth\else\Gin@nat@width\fi}
\def\maxheight{\ifdim\Gin@nat@height>\textheight\textheight\else\Gin@nat@height\fi}
\makeatother
\setkeys{Gin}{width=\maxwidth,height=\maxheight,keepaspectratio}

\usepackage[font={small},skip=2pt]{caption}
\usepackage{float}
\hypersetup{breaklinks=true,
            bookmarks=true,
            pdfauthor={},
            pdftitle={VIF: Virtual Interactive Fiction (with a twist)},
            colorlinks=true,
            citecolor=black,
            urlcolor=blue,
            linkcolor=black,
            pdfborder={0 0 0}}
\usepackage{balance}

\title{VIF: Virtual Interactive Fiction (with a twist)}

\author{
     \alignauthor{%
    \textbf{Jérémy Frey}\\
    \affaddr{Univ. Bordeaux} \\
    \affaddr{351 Cours de la Libération} \\
    \affaddr{33400 Talence, France} \\
    \email{jeremy.frey@inria.fr}\\
 } 
   }

\date{}

\usepackage{balance}

\usepackage{comment}

\usepackage[scaled=.92]{helvet} 

\usepackage{suffix}

\usepackage{marginnote}
\reversemarginpar

\begin{document}

      \teaser{
      \includegraphics[width=\textwidth]{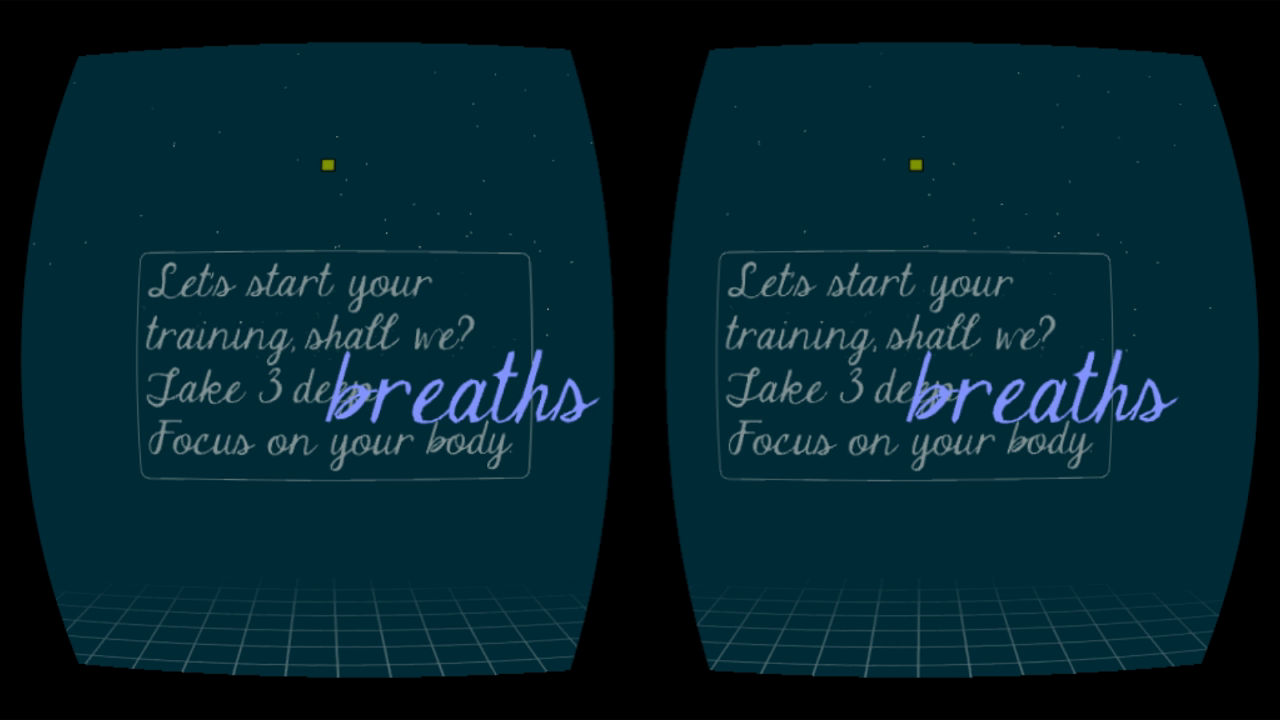}
      \caption{View of the virtual environment. Part of the text is dynamically mapped
to player's physiology, such signals also influence the narrative.\label{fig:teaser}}
    }
  
\maketitle

\RaggedRight{} 

\begin{abstract}
Nowadays computer science can create digital worlds that deeply immerse
users; it can also process in real time brain activity to infer their
inner states. What marvels can we achieve with such technologies? Go
back to displaying text. And unfold a story that follows and molds users
as never before.
\end{abstract}



\def \citep {\protect\cite}

\WithSuffix\newcommand\caption*{\caption}

\reversemarginpar
\marginnote{
\begin{minipage}{\marginparwidth}
   \vspace{-19ex}
   \includegraphics[width=1\marginparwidth]{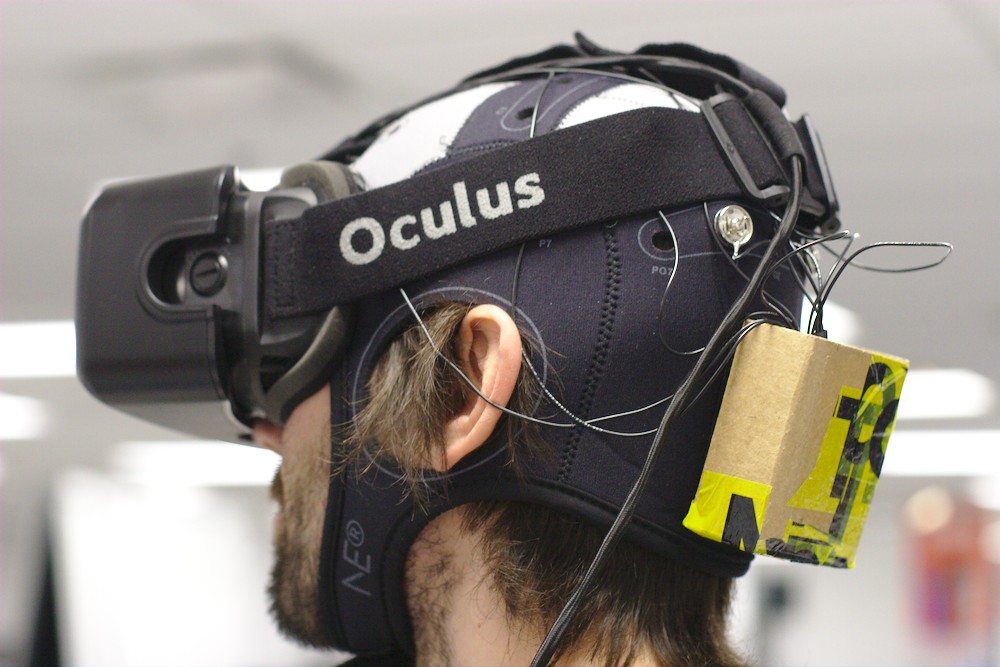}
   \captionof{figure}{The Oculus Rift DK1 HMD\protect\footnote{\url{https://www.oculus.com/}} is combined with Neuroelecronics Enobio EEG cap\protect\footnote{\url{http://www.neuroelectrics.com/}} and OpenBCI amplifier\protect\footnote{\url{http://www.openbci.com/}} to both immerse and monitor user's brain activity. 
    \label{fig:hmdeeg}
   }
\end{minipage}
}

\section{Introduction}\label{introduction}

On the one hand, we have devices that are more than ever capable to
create a variety of illusions and to draw attention up to the point that
users will forget about the real world -- the notion of virtual presence
\citep{Slater2009}. The paragons of such devices are probably at this
moment head-mounted displays (HMD), that, \emph{literally}, replace what
people see with a virtual world -- so called virtual reality (VR).
Combined with headphones, the reality gets obliterated for
most\footnote{Even though the focus is here on textual stories with
  (optional) ambient sounds, any other combination of input modalities
  -- e.g.~narrate a story and display (optional) colors -- is of course
  suitable.}.

On the other hand, physiological computing is mature enough to asses
mental states and emotions \citep{Fairclough2009a}. In particular,
mobile brain imaging techniques alike electroencephalography (EEG) can
be employed to measure a variety of constructs such as workload or
attention; cognitive processes that were harder to monitor in the past
\citep{Frey2014a}. Physiological sensors are not restricted to the
laboratory, nowadays they can be used the field through carefully
crafted wearables, e.g. \citep{Gervais2016}. These technologies make
computers \emph{comprehend} users \citep{Picard1995}.

On the \emph{other} other hand, is lying interactive fiction (IF),
stories that depend on the actions of the readers/players\footnote{In
  the present document I consider a broad definition of IF, that
  encompasses both Choose-Your-Own-Adventure (CYOA) games and games
  where users have to actually type text to interact.}, a genre that is
gaining increased attention from outside its original community
\citep{Montfort2012}. The form of IF evolved from choice-based adventure
games to complex generative systems that can narrate a living world
\citep{Montfort2011}. Besides advances in natural language processing,
IF attracted new game-makers thanks to writing tools -- based on markup
languages or graphical programming -- easier to handle\footnote{One
  could find a variety of commented IF works on
  \url{https://emshort.wordpress.com/}.}.

\reversemarginpar
\marginnote{
\begin{minipage}{\marginparwidth}
   \vspace{-80ex}
   \includegraphics[width=1\marginparwidth]{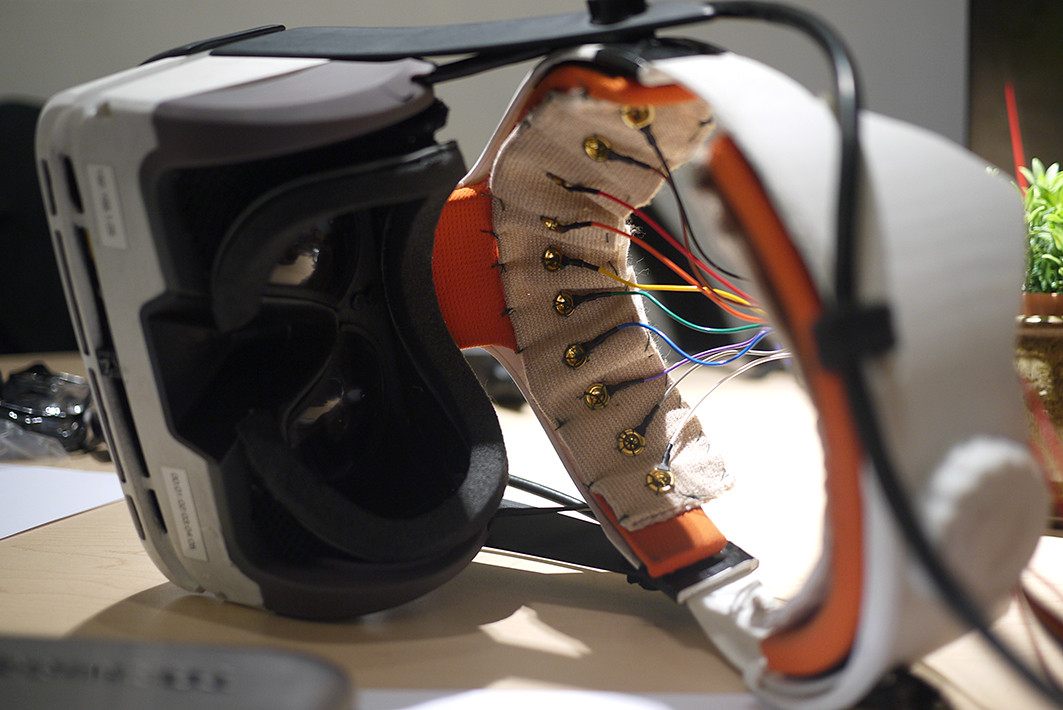}
   \captionof{figure}{Prototype of a HMD device that directly embeds EEG electrodes -- here a Vrvana Totem\protect\footnote{\url{http://www.vrvana.com/}}. Picture © Alexandre Girardeau.
    \label{fig:totem}
   }
\end{minipage}
}

\reversemarginpar
\marginnote{
\begin{minipage}{\marginparwidth}
   \vspace{-20ex}
   \includegraphics[width=1\marginparwidth]{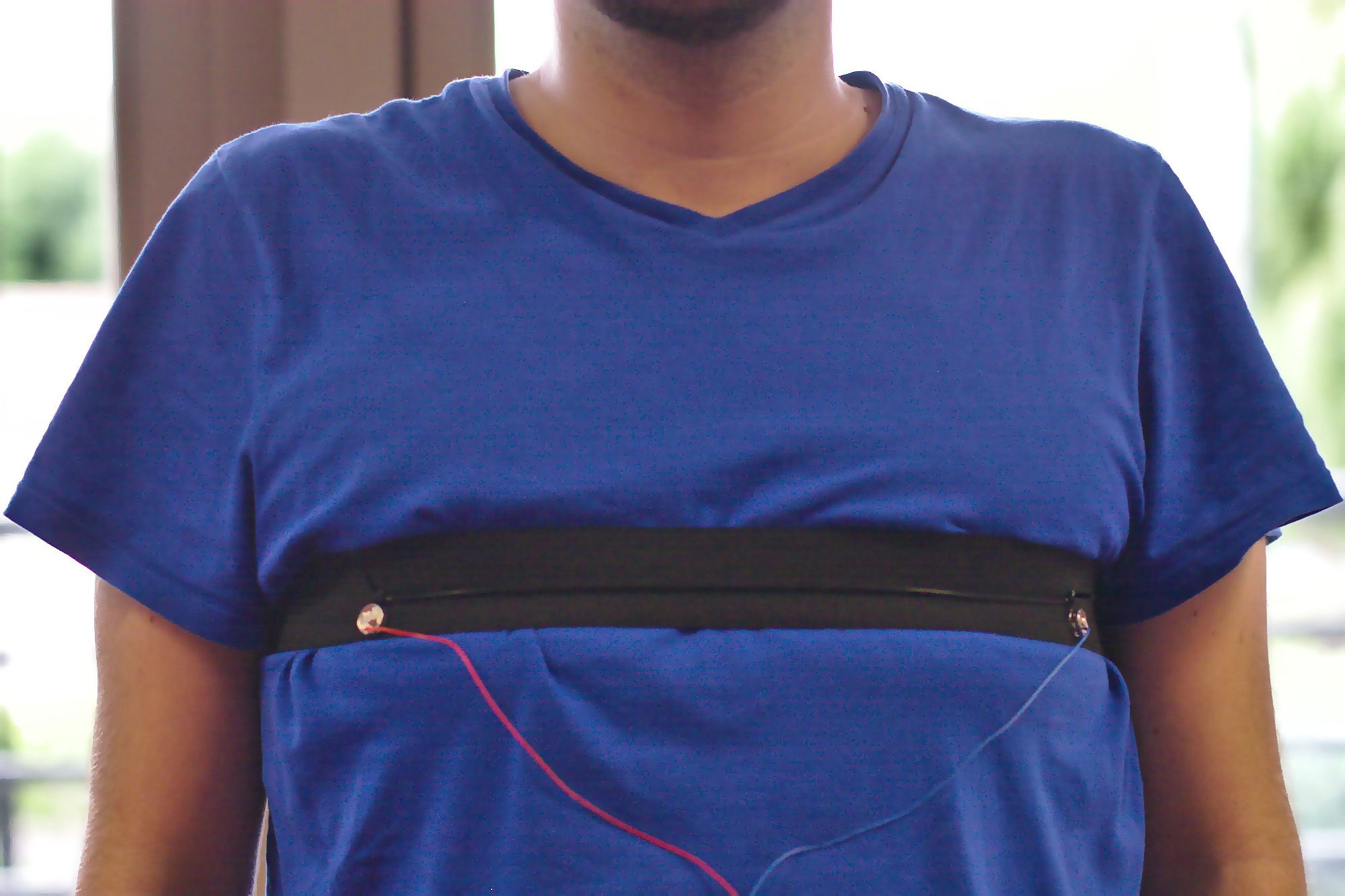}
   \captionof{figure}{Various sensors could be integrated, e.g. a belt that measures breathing. From \citep{Gervais2016}.
    \label{fig:breathing}
   }
\end{minipage}
}

\reversemarginpar
\marginnote{
\begin{minipage}{\marginparwidth}
   \vspace{10ex}
   \includegraphics[width=1\marginparwidth]{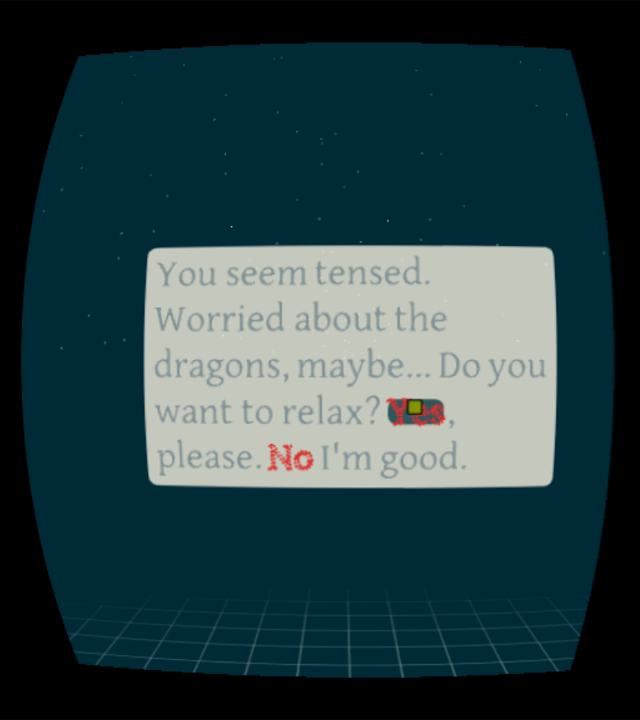}
   \captionof{figure}{Users can interact with text elements in a more traditional way, using their gaze. 
    \label{fig:select}
   }
\end{minipage}
}

Putting hands back together, physiological sensors and brain computer
interfaces could be added to the repertoire of IF makers in order to
craft more resonating stories, that can both adapt and react to players.
A narrative branch could be generated based on the (past) affect of the
user, in order to favor or avoid an emotion. The events or the phrasing
could be driven by the attentional state of the user to increase
comprehension and seek a state of flow. Even more involving and
pervasive, the user could be asked to perform breathing exercises and
get actually relaxed before starting an adventure in a gloomy dungeon --
a journey of initiation for the wannabe warrior that will echo within
the reader.

While rich inputs derived from physiology are already investigated with
traditional video games \citep{Nijholt2009, Nacke2011}, the ``form
factor'' of IF should ease the integration of these modalities to the
fabric of the story. Indeed, readers are used to filling the gaps
between ellipses or suggested details; through their imagination they
seize the content and make it their own. To change the mood of a scene,
whereas a AAA video game will have to smoothly adapt assets that where
expensive to build in the first place, a writer will lightly tune the
right adjective. Simple and seamless.

Far from being a trendy gimmick, the use of HMD and VR let readers focus
entirely on the text, preventing stimuli external to the narrative to
reach them and distract them from the story\footnote{Resulting in an
  \emph{immersive} book indeed.}. The simplicity of the scene will
hopefully reduce perceptual load, freeing more cognitive resources to
let them build their own representations, increasing engagement. At the
same time, HMD could be used to design simple game mechanisms based on
gaze and head tracking, yet another input modality for readers to
\emph{physically} interact with IF (see System Description below).

First and foremost, this technology is really a mean to put upfront one
medium that shaped for ages humanity \citep{Bettelheim1976}. If the word
is mightier than the tools, now it shall be exalted by computer science
to help twist our minds for the better.

\section{Description of the system}\label{description-of-the-system}

This section describes a proof of concept that implements a basic set of
features to support the ideas discussed previously. The program and its
source code are available on \url{https://github.com/jfrey-xx/vif}.

\subsection{Rendering}\label{rendering}

The graphics are kept simple on purpose. As an emanation of IF -- even
in 3D and in stereoscopy -- the virtual world is mostly composed of text
blocks. Generative typography and colors can impersonate the narrator or
characters. They can also be used to give a biofeedback -- e.g. ``your
heart \emph{beat}'' could pulse toward the camera and change color along
each heartbeat (Figure \ref{fig:teaser}) -- depending on the control or
awareness given to users (see also \citep{Gervais2016}).

Other graphical elements are present, but only to fulfill particular
functions. A grid on the floor gives a sense of position in space. The
day/night cycle has two purposes: to give a sense of orientation (the
trajectory of the sun gives cardinal directions) and to give a sense of
time (Figure \ref{fig:time}).

\subsection{Interaction}\label{interaction}

The main contribution comes from the events that can be triggered
depending on the physiology and on the mental state of the players --
e.g.~three deep breaths, getting relaxed, \ldots{} However, alike
choice-based adventures, players can also select explicitly interactive
chunks of words -- e.g. ``open the box'' / ``feed the box to my pony''.
They use their gaze to do so (Figure \ref{fig:select}).

Time passing by can be measured to drive the narration and the text
could change accordingly to the time of the day -- e.g.~wolves are
obviously out at night.

The virtual world surrounding players, the text can be positioned at
various places, both related to the orientation of the player or to
cardinal directions (e.g. ``behind'', ``North''). Then the field of view
of the player can be taken into account in order to show or hide text
blocks based on what is seen (Figure \ref{fig:place}). Sentences that
could creep from behind: ideal for an horror novel.

\subsection{Sensors}\label{sensors}

Wearables such as belts measuring breathing (Figure \ref{fig:breathing})
or smartwatches measuring heart rate can be plugged to the system.
However, since players are already equipped with HMD, this latter piece
of hardware can be used at the same time to embed sensors. For instance,
thanks to affordable hardware coming from the DIY movement, a (very)
early prototype of a HMD with integrated EEG electrodes was crafted to
measure brain activity (Figure \ref{fig:totem}). Regular EEG caps could
also be combined with HMD (Figure \ref{fig:hmdeeg}). Eventually, a whole
array of physiological measures may be inferred directly from a device
placed on the head -- see, e.g. \citep{Hernandez2015}.

\reversemarginpar
\marginnote{
\begin{minipage}{\marginparwidth}
   \vspace{-45ex}
   \includegraphics[width=1\marginparwidth]{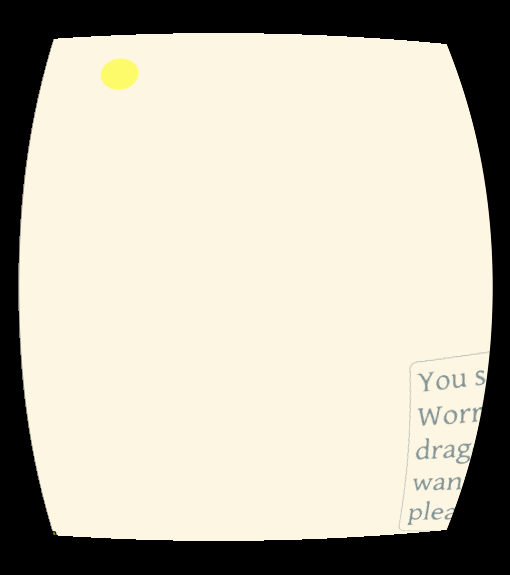}
   \includegraphics[width=1\marginparwidth]{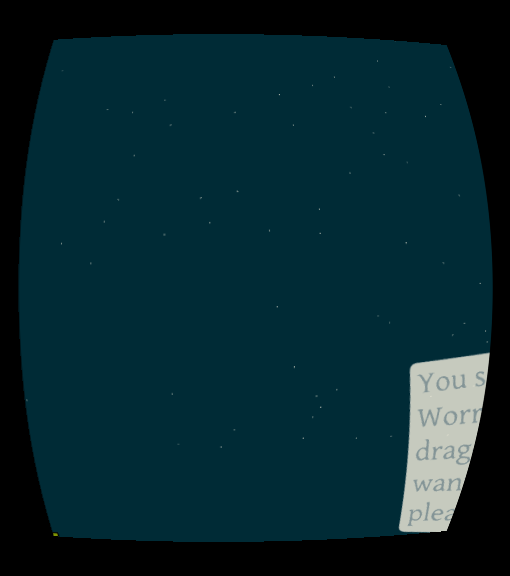}
   \captionof{figure}{Besides text, graphical elements are kept minimal. The environment serves purposes, e.g. to give a sense of time or a sense of orientation. 
    \label{fig:time}
   }
\end{minipage}
}

\marginnote{
\begin{minipage}{\marginparwidth}
   \vspace{30ex}
   \includegraphics[width=\marginparwidth]{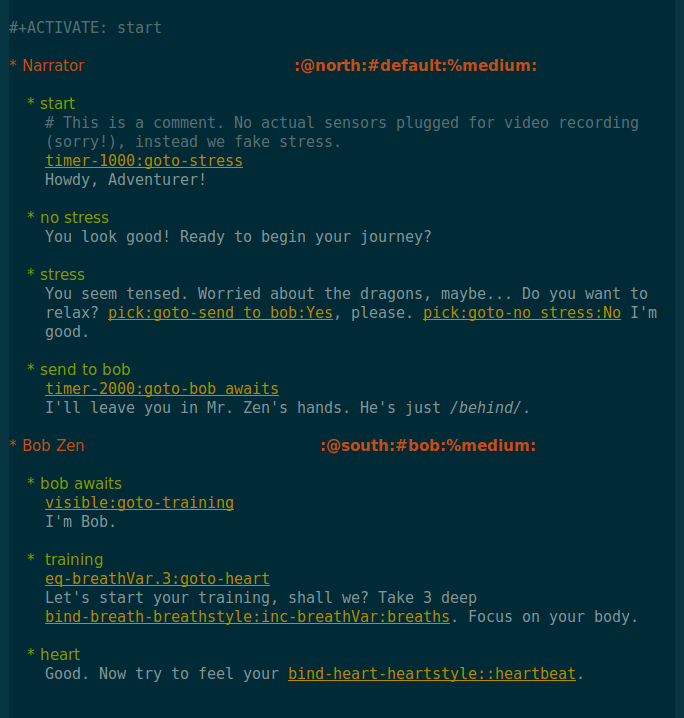}
   \captionof{figure}{Game-makers / writers can use their favorite text editor and a markup language to create new content. 
    \label{fig:parser}}
\end{minipage}
}

\subsection{Writing 101}\label{writing-101}

Game-makers with a background in programming can suit the engine behind
VIF to their needs, for example add new typographic effects. However,
for regular users that simply wish to create new content, a text parser
based on a markup language has been implemented (Figure
\ref{fig:parser}).

\subsection{Beyond: going social}\label{beyond-going-social}

Since communications between the game engine and physiological measures
rely on a network protocol, it would be trivial to share and sync data
between multiple instances of the system. Doing so would extend the
system with multiplayer elements. Several in-game characters could then
be personified by others users -- the states and choices of everyone
influencing the overall story --; direct or intimate interactions that
could deeply connect players, help them share an experience.

\section{Inspirations}\label{inspirations}

The idea of using processed physiological signals as a pervasive element
roots to the affective computing movement \citep{Picard1995}, even
though since these debuts the spectrum of available measures has widened
beyond emotions through physiological computing \citep{Fairclough2009a}
and neuroadaptive technologies \citep{Zander2011}.

My personal (re)discovery of the IF genre comes from the productions of
the Inkle Studios\footnote{\url{http://www.inklestudios.com/}}. Whereas
I could not achieve a CYOA book when I was a child, their clever games
designs kept me engaged throughout the stories, giving a clue on how to
hook readers. Later on I found interesting perspectives on the whys and
hows we may rely so much on stories within the preliminary chapters of
\citep{Bettelheim1976}, Bruno Bettelheim's book describing how
significant fairy tales could be during child development.

However, these cornerstones are a mere \emph{retcon} of what aroused my
interest for interactive and adaptive fictions in the first place. All
started as it should, with a story within a story. In Ender's
Game\citep{Card1985}, Orson Scott Card describes how a RPG-like video
game used by the military to monitor young recruits' mental health is
``hijacked'' by an alien race in order to reach out for the hero and
plead for empathy\footnote{Note that novel's background is itself a
  (too) serious game.}.

\reversemarginpar
\marginnote{
\begin{minipage}{\marginparwidth}
   \vspace{-40ex}
   \includegraphics[width=1\marginparwidth]{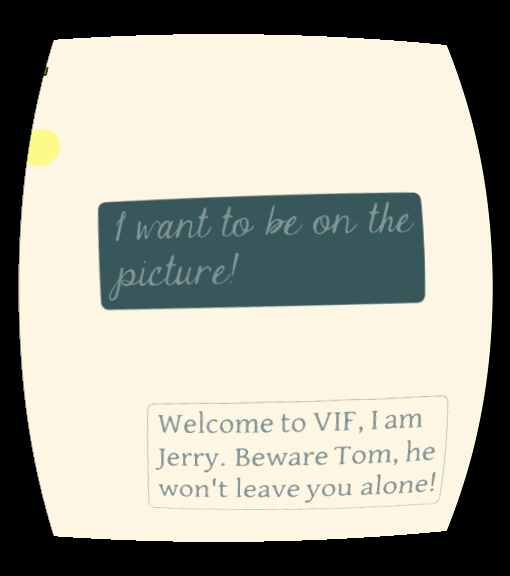}
   \includegraphics[width=1\marginparwidth]{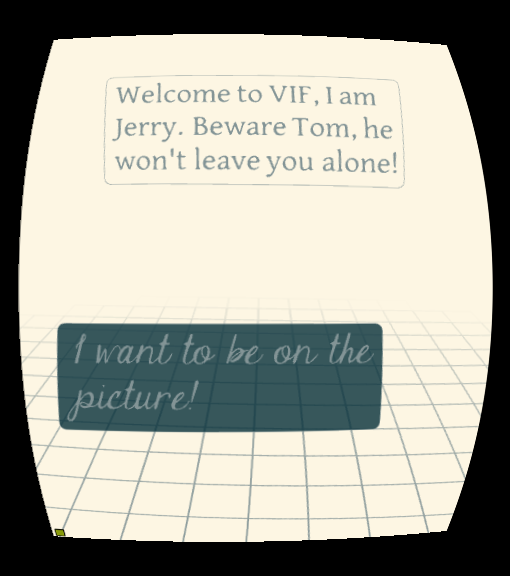}
   \captionof{figure}{Events can occur depending on what players see and on texts' position, resulting in emerging game mechanisms that are physically involving.
    \label{fig:place}
   }
\end{minipage}
}

\section{Superpowers}\label{superpowers}

As per workshop requirement, few lines about my abilities. During my PhD
I started to explore how physiological computing can contribute to
human-computer interaction and foster new communication channels among
people. In particular, I have been using EEG and machine learning to
measure constructs such as workload, comfort or attention -- dimensions
of the user experience that I was able to grasp thanks to my background
in cognitive science. With hacking spells, I helped to craft sensors
more practical to use in the field, an opportunity to investigate
multi-users scenarios and playful contexts. I also participated in the
development of novel physiological feedback, that could be integrated
seamlessly to the physical surroundings through spatial augmented
reality and tangible user interfaces. Science-fiction reader at night to
fetch ideas, daydreamer to create new ones, I am not afraid to get my
hands dirty to implement them, more eager to do so when they sound crazy
or stupid, even when I have already too many projects running for my own
good. In the end I seek to use computer science as a mean to enhance
well-being and facilitate human relationships on the whole. Mandatory
website: \url{http://phd.jfrey.info/}.

\section{Acknowledgements}\label{acknowledgements}

After months toying with the idea, I finally had the opportunity (and
the liberty!) to implement a first version of this project while I was
visiting the MuSAE Lab in late 2015, under the kind supervision of Tiago
H. Falk. During my stay in Montreal, the HMD+EEG cocktail would never
have come to life without the vivid Breathe@Work crew that gathered
during the Hacking Health event, and for \emph{that} to happen, I was
lucky to meet two wonderful communities there, namely NeuroTechX and
Highway 101. Since VIF is but a spin-off of the Tobe toolkit, I shall
not to forget my usual co-authors \citep{Gervais2016}. Finally, Joan Sol
Roo is the one who made me realize the connection with Ender's Game,
that explained it all.

\balance{}

\bibliography{biblio}

\end{document}